\documentclass[aps,prb,showpacs,twocolumn,superscriptaddress,floatfix,letterpaper]{revtex4}

\usepackage[dvips]{graphicx}

\def\ee#1{\times 10^{#1}}
\def\vec#1{\mathbf{#1}}
\def\unit#1{\mathrm{#1}}

\begin{document}



\title{Vortex state oscillations in soft magnetic cylindrical dots}



\author{K.~Yu.~Guslienko}
\email{gusliyenko@anl.gov}
\affiliation{Materials Science Division, Argonne National Laboratory,
9700 S.~Cass Ave., Argonne, IL 60439}

\author{W.~Scholz}
\affiliation{Seagate Research, 1251 Waterfront Place, Pittsburgh, PA 15222}

\author{R.~W.~Chantrell}
\affiliation{Seagate Research, 1251 Waterfront Place, Pittsburgh, PA 15222}

\author{V.~Novosad}
\affiliation{Materials Science Division, Argonne National Laboratory,
9700 S.~Cass Ave., Argonne, IL 60439}


\date{\today}

\begin{abstract}

We have studied magnetic vortex oscillations in soft sub-micron cylindrical dots
with variable thickness and diameter by an analytical approach and 
micromagnetic simulations.
We have considered two kinds of modes of the vortex magnetization oscillations: 
1)~low-frequency
translation mode, corresponding to the movement of the vortex as a whole near 
its equilibrium
 position; 2)~high-frequency vortex modes, which correspond to radially 
symmetric oscillations
 of the vortex magnetization, mainly outside the vortex core.
The vortex translational eigenmode was calculated numerically in frequency and 
time domains for different dot aspect ratios.  
To describe the discrete set of vortex high-frequency modes we applied the 
linearized equation of
motion of dynamic magnetization over the vortex ground state. We 
considered
only radially symmetric magnetization oscillations modes. The eigenfrequencies 
of both kinds of excitation modes are determined by magnetostatic interactions. 
They are proportional to the thickness/diameter ratio and lie in the GHz range 
for typical dot sizes. 


\end{abstract}




\maketitle


\section{Introduction}

The magnetization vector of a ferromagnetic body (particle) oscillates when 
variable external magnetic fields are applied. The amplitude of magnetization 
oscillations, which depends on the intrinsic fields as well as on the particle's
geometrical parameters increases drastically when the frequency of the external 
field coincides with the frequency of a fundamental spin-excitation eigenmode 
of the system. 

This issue has important implications for developing high-speed and 
high-density magnetic devices, where short field pulses are applied to magnetic 
media to achieve the magnetization reversal.\cite{lyberatos02} The 
characteristic switching time is usually in the nanosecond and sub-nanosecond 
range for non-uniform and coherent magnetization rotation, respectively.
Understanding of spin-wave excitation spectra is thereby crucial to determine 
the field-dependent spin-instability regions, where spontaneous or 
thermally-assisted magnetization reversal processes might occur 
\cite{chubykalo02}.

In the general case, both the short-range exchange and the long-range 
magnetostatic interactions contribute to eigenfrequencies of the collective 
spin excitations.
The long-wavelength excitations are mainly influenced by dipole-dipole 
interactions. Therefore, the demagnetizing fields of the magnetic elements 
determine their eigenfrequencies. These fields can be strongly non-uniform in 
the case of non-ellipsoidal samples or if the static magnetization distribution 
differs significantly from the single-domain (collinear) state.
This is especially 
the case for technologically important flat magnetic particles, prepared by 
means of micro-fabrication and thin film deposition techniques. The 
adequate description of high-frequency magnetization dynamics in such systems 
is a challenge for modern magnetism theory.

The remanent magnetization distribution within a ferromagnetic particle is 
a complex
phenomenon that depends on the size and shape of the particle as well as on the 
balance of different contributions (exchange, magnetostatic, anisotropy etc.) 
to the magnetic energy. Exchange energy dominates at small particle sizes and 
favors a uniform magnetization distribution, so-called single-domain or 
``flower'' (``leaf'') state. On the other hand, this almost collinear spin 
alignment leads
to large demagnetizing fields due to magnetic surface charges and 
correspondingly increase of the magnetostatic energy.  The importance of the 
magnetostatic energy increases gradually as the particle size increases. The competition 
between these two energies and anisotropy energy leads to non-uniform magnetic 
states such as multi-domain states (if the anisotropy constant is greater than 
the magnetostatic energy) or magnetic
vortices (if the anisotropy energy is small in comparison with the 
magnetostatic energy).
The critical size at which a particle becomes single domain depends on the 
interplay between the energies mentioned above, and also on the shape of the 
particle. 
 This critical size is of the order of the material exchange length 
$L_\mathrm{ex}=\sqrt{2A/M_\mathrm{s}^2}$ , where $A$ is the exchange
stiffness constant and $M_\mathrm{s}$ is the saturation magnetization. A 
magnetically soft particle with lateral dimensions smaller than the exchange 
length ($L_\mathrm{ex} \approx 10-20$~nm) is expected to remain in a single 
domain state at remanence.
  
   Sub-micron magnetic elements with regular shapes (rectangular, cylindrical, 
spherical etc.) have attracted much attention during the last few years due to 
considerable progress in fabrication and characterization of the 
samples.\cite{ross01, ross02}

In this paper we will consider dynamic properties of circular particles (dots)
made of soft magnetic material with a thickness of about $L_\mathrm{ex}$ 
and diameters considerably larger than $L_\mathrm{ex}$. It is well established, 
that a magnetic vortex structure is the ground state of such submicron 
dots.\cite{shinjo00, raabe00, schneider00, cowburn99, novosad01, novosad02, 
guslienko02_2} This vortex state can be interpreted as a part of a 
two-dimensional magnetic
topological soliton\cite{kosevich90} adapted to the particle shape to minimize 
the total magnetic energy. 
Experimental studies of the vortex
state in flat submicron dots have been performed using different methods. 
Magnetization reversal in sub-micron dots due to magnetic vortex formation and 
its displacement has been explored by analytical
and numerical micromagnetic modeling.\cite{guslienko01,guslienko01_2}
It was found from static measurements and calculations
that the magnetic vortex state corresponds to a deep energy minimum for 
submicron dot
radii and it can be destabilized either by reducing the dot radius down to the 
material's exchange
length or by applying an in-plane magnetic field.
The vortex state is not specific for the circular shape, it was also observed 
in rectangular,\cite{cowburn00} elliptical,\cite{fernandez00, grimsditch98} and 
triangular\cite{koltsov00} flat magnetic elements.
  
  The dynamic properties of sub-micron dots have only more recently attracted 
some attention.\cite{novosad02}
There have been a few successful experimental investigations of the
magnetization dynamics under short field pulses in a saturated FeNi 
disk\cite{hiebert97} and also
closure-domains in a Co disk.\cite{acremann00}
In this article we concentrate on the linear spin dynamics in an equilibrium, 
but essentially
 non-uniform, vortex state. Spin excitations in the vortex-state are expected 
to be substantially different from those in
the uniformly magnetized state and also from the excitations (spin-waves) 
observed in infinite thin
magnetic films.  An example is the appearance of a low-frequency mode 
(translation mode)
associated with  the movement of the vortex as a whole in a potential well
created mainly by magnetostatic energy.\cite{guslienko02,ivanov04}
It was found that the vortex core undergoes a spiral motion with the 
eigenfrequency in the sub-GHz range for thin sub-micron dots. This lowest 
frequency translation mode can be treated as vortex core excitation and is 
localized near the
dot center.  This mode is similar to crossed-domain wall resonance in a 
magnetic film.\cite{argyle84}
We consider that the spectrum of vortex spin excitations will be quantized due 
to particle's
(dot's) restricted geometry as was shown in\cite{guslienko00,jorzick01} for the 
saturated state.
Spin excitations of the vortex state have been discussed for an infinite film 
(Ref.~\onlinecite{ivanov98} and Refs.\ therein) 
neglecting the magnetostatic interaction.  This is an unsuitable approximation
to describe magnetization dynamics of sub-micron magnetic particles, where the 
exchange contribution
 to the low-lying eigenfrequencies is essentially smaller than the 
magnetostatic contribution ($L_\mathrm{ex} \ll R$, where $R$ is the disk 
radius).
Magnetization dynamics in this case is governed by the long-range dipolar 
forces.  We are not aware of any
theoretical or experimental investigation of dipole dominated high frequency 
spin excitations in a
magnetic vortex state except recent results by Novosad et al.\cite{novosad02} 
and Ivanov et al.\cite{ivanov04}

   In this paper we address the precessional spin dynamics of the vortex ground 
state and determine the low-lying vortex eigenfrequencies which correspond to 
radially symmetric spin excitations. The eigenfrequencies are well above the 
vortex translation eigenfrequency. Analytical calculations of the 
eigenfrequencies are supported by numerical micromagnetic simulations. 

The vortex magnetization dynamics can be detected, for instance, by 
ferromagnetic resonance (FMR), Brillouin light scattering (BLS) techniques or by 
time-resolved Kerr microscopy.  The spectrum of the dot discrete 
eigenfrequencies
can be used as a starting point for the consideration of magnetization reversal.

We present in Sec.~\ref{s_theory} a theoretical description and in
Sec.~\ref{s_mumag} micromagnetic calculations of discrete
eigenfrequencies of excitation modes of the vortex state in soft magnetic cylindrical dots. 
We consider that
above the vortex translation mode frequency (which is $<1$~GHz)
there is a set of 
radially and azimuthal symmetric modes
(of about 10~GHz) localized outside the vortex core. For sub-micron dot sizes 
the dynamic demagnetizing
 fields determine the eigenfrequencies of these modes, whereas the translation 
mode
 frequency is determined by the static demagnetizing field (magnetostatic 
contribution to the 
restoring force). The vortex excitation eigenfrequencies are found from 
eigenvalues of a magnetostatic integral operator (Sec.~\ref{s_theory})
and also from dynamical micromagnetic 
simulations (Sec.~\ref{s_mumag}) based on the
Landau-Lifshitz equation of motion. Full details of the analytical model are 
reported. Experimental techniques to study these vortex modes are discussed.
A comparison of numerical (FEM-BEM and finite differences) micromagnetic 
simulations with analytical results is presented. Finally, the summary is given 
in Sec.~\ref{s_conclusions}.

\section{\label{s_theory}Theory}

We present a theoretical description of the vortex state oscillations in 
submicron-size cylindrical dots. Our approach is based on the consideration of 
small dynamic oscillations over the centered vortex
 ground state. We consider here only radially symmetric vortex excitation modes 
in zero applied magnetic field. These
modes have frequencies well above the vortex core translation frequency for the 
given dot sizes. They have maximum FMR and BLS intensity due to their non-zero 
average dynamic magnetization.
It is known that the vortex core bears some topological charges (gyrovector and 
vorticity, see Ref.~\onlinecite{kosevich90}) and this peculiarity has important consequences for the vortex state 
dynamics. In the following, we do not consider the vortex charges
and assume that the high-frequency 
excitation modes are localized outside the vortex core. In other words, 
we neglect the vortex core influence on these modes. 
Let $L$ denote the dot thickness, and $R$ the dot radius.
 Since the dot thickness is small, we assume a two-dimensional magnetization 
distribution
 $\vec{m}=\vec{M}/M_\mathrm{s}$, $\vec{m}^2=1$
within the cylindrical dot, which does not depend on the $z$-coordinate along the dot 
thickness. We use
the angular parameterization for the dot magnetization components 
$m_z=\cos\Theta$,
$m_x+im_y=\exp(i\Phi)\sin\Theta$, $\Phi=\phi \pm \pi/2$  .
Here positive and negative signs correspond to counter-clockwise and
clockwise rotation of the vector 
$\vec{m}$ in the dot plane, respectively. The cylindrical coordinates $\rho, 
\phi, z$ are used below.
The spin structure of the static vortex located in the dot center may be 
described by the one-parameter ansatz
\begin{equation}\label{Eq0}
\tan(\Theta_0(\rho)/2)=\rho /b
\label{eq_theta0}
\end{equation}
if $\rho<b$  and  $\Theta_0=\pi/2$ if $\rho>b$ as suggested by Usov et 
al.\cite{usov93} The radius of the vortex ``core'' $b \approx 10$~nm
can be determined from magnetic energy minimization. 
The total dot magnetic energy including exchange and magnetostatic energies is:

\begin{equation}\label{Eq1}
W= L \int{d^{2}\vec{\rho}
[A(\nabla\Theta)^{2}+\sin^{2}\Theta(\nabla\Phi)^{2})-\frac{1}{2}\vec{H}_{m}\cdot
\vec{M}]},
\end{equation}
where $\vec{H}_m$ is the magnetostatic field. We use the equation of motion of 
variable magnetization

\begin{equation}\label{Eq2}
-\frac{1}{\gamma} \frac{d\vec{m}}{dt}=\vec{m}\times(-\frac{\delta
W}{\delta\vec{m}}) \quad.
\end{equation}

The calculation of the exchange contribution to the effective field is
straightforward, while the magnetostatic field is expressed via the
magnetostatic Green's function\cite{guslienko00} in cylindrical coordinates 
$\rho, \phi,
z$. The magnetization distribution $\vec{m}(\vec{r})$ within the cylindrical dot
leads to the variable demagnetizing field 
$\vec{h}(\vec{r})=\widehat{G}[\vec{m}(\vec{r})]$, where  $\widehat{G}$ is a 
tensorial
non-local integral operator (the tensorial magnetostatic Green's
function $G_{\alpha\beta}(\vec{r},\vec{r'})$) expressed in the cylindrical 
coordinates

\begin{eqnarray}\label{Eq3}
\widehat{G}[\vec{m}(\vec{r})] & = & \int 
\widehat{G}[\vec{r},\vec{r'}]\vec{m}(\vec{r'})d^{3}\vec{r'}, \nonumber \\
G_{\alpha\beta}(\vec{r},\vec{r'})
& = & 
-(\nabla_{\vec{r}})_{\alpha}(\nabla_{\vec{r'}})_{\beta}\frac{1}{\mid\vec{r}-\vec
{r'}\mid}
\end{eqnarray}

 We use averaging over $z, z'$ and also over $\phi'$ (due to the assumed radial 
symmetry of $\vec{m}(\vec{r})$).
 Only $\rho \rho$ and $zz$-components of the averaged $\widehat{G}$ tensor  are 
not equal to zero. The equation of motion (\ref{Eq2}) is then linearized
by substituting  
$\vec{m}(\vec{\rho},t)=\vec{m}_{0}(\vec{\rho})+\vec{\mu}(\vec{\rho},t)$  or by 
magnetization angles:

\begin{equation}\label{Eq4}
\Theta(\rho,t)=\Theta_{0}(\rho)+\vartheta(\rho,t),  \Phi(\rho,\varphi,t)= 
\varphi+\frac{\pi}{2}+\psi(\rho,t),
\end{equation}
where the vector 
$\vec{\mu}=(\mu_{\rho},\mu_{\varphi},\mu_{z})=(-\psi\sin\Theta_{0},\vartheta\cos
{\Theta_{0}},-\vartheta\sin{\Theta_{0}})$
is the dynamic part of the magnetization. For small variables $\vartheta$
and $\mu =\mu_{\rho}=-\psi\sin\Theta_{0}$ we get a system of two coupled 
integro-differential equations. We can neglect the dynamic exchange interaction 
for sub-micron dot radii, exclude the variable $\vartheta (\rho ,t)$ and reduce 
the problem to an eigenvalue problem for the integral magnetostatic operator:

\begin{equation}\label{Eq5}
(\frac{\omega}{\gamma
M_{s}})^{2}\mu(\rho) = \int{d\rho'\rho'\Gamma(\rho,\rho')\mu(\rho')},
\end{equation}

\begin{equation}\label{Eq6}
\Gamma(\rho,\rho')=
\int{dr \, r\sin\Theta_{0}(r)g_{zz}(\rho,r)g_{\rho\rho}(r,\rho')},
\end{equation}
where the averaged over thickness Green's function components do not depend on 
$\phi$:

\begin{equation}\label{Eq7}
g_{\alpha\beta}(\rho,\rho^{'})= \frac{1}{L}
\int_{0}^{L}{dz\int_{0}^{L}{dz'\int_{0}^{2\pi}d\phi'G_{\alpha\beta}(\vec{r},\vec
{r'})}}
\end{equation}

Solution of Eq.~(\ref{Eq5}) yields the eigenfunctions $\mu _n(\rho)$
and corresponding 
eigenfrequencies $\omega_n$, which can be numbered by integers $n=1,2,\ldots$ 
omitting the azimuthal index $m=0$ (radial symmetry).
We use the index $n=0$ for the 
vortex translation mode.\cite{guslienko02} This mode has azimuthal indices $m=\pm1$. We do not 
assume any boundary conditions for the dynamic magnetization on the dot side 
boundary due to negligibly small surface anisotropy for soft magnetic materials 
(FeNi, for instance). The vortex core
is a stable formation which is strongly coupled by exchange forces and we 
assume that the dynamic magnetization has considerable values only outside of 
the vortex core, in the area where $\Theta_{0}(\rho)=\pi/2$ .
 Then the dynamic magnetization will have only two components, $\mu_{\rho}$ and 
$\mu_z$ given by Eq.~(\ref{Eq0}),
which are perpendicular to the vortex static magnetization  
$\vec{m}_0=(0,m_0^{\phi},0)$.
These components depend on time as $\mu_{\rho}(t) \sim \sin(\omega t+\phi_0)$, 
$\mu_{z}(t) \sim \cos(\omega t+\phi_0)$,
 where $\phi_0$ is an initial phase. This time dependence corresponds to 
elliptical precession
of the dynamic magnetization components in the $\rho-z$ plane around the 
vortex static equilibrium
magnetization $\vec{m}_0$.
The dynamic $\mu_{\rho}$  component determines the dynamic side surface and 
volume magnetic charges. The coupled oscillations of $\mu_{\rho}$ and $\mu_{z}$ then 
lead to a
considerable increase of the dynamic magnetostatic energy (mainly due
to volume charges, because surface charges described by $\mu_{\rho}(R)$ 
magnetization component are small) and to more high eigenfrequencies in 
comparison with the case of the vortex translation mode (no induced
surface charges). Assuming that the dynamic components are concentrated mainly 
outside of the vortex core, in the area $\Theta_{0}(\rho)=\pi/2$, for small 
dot aspect ratio $L/R\ll1$ (in this case
$g_{zz}(\rho,\rho')=-4\pi\delta(\rho-\rho')/\rho$) we get from
Eqs.~(\ref{Eq6}) and (\ref{Eq7}) the simplified integral equation

\begin{equation}\label{Eq8}
(\frac{\omega}{\omega_{M}})^2 \mu(\rho)= -
\frac{1}{4\pi}\int_{0}^{R}d\rho'\rho'g_{\rho\rho}(\rho,\rho')\mu(\rho'),
\end{equation}

\begin{equation}\label{Eq9}
g_{\rho\rho}(\rho,\rho')=-4\pi\int_{0}^{\infty}dkkf(kL)J_{1}(k\rho)J_{1}(k\rho')
,
\end{equation}

\begin{displaymath}
f(x)=1-(1-e^{-x})/x \quad , \quad
\omega_M=\gamma 4\pi M_\mathrm{s}.
\end{displaymath}
The eigenfrequency $\omega _n$ which corresponds to the normalized to unit 
n-eigenfunction $\mu_n(\rho)$ is
\begin{equation}\label{Eq_wn_wM}
(\frac{\omega_{n}}{\omega_{M}})^2=
-\frac{1}{4\pi}\int_{0}^{R}d\rho\rho
\int_{0}^{R}d\rho'\rho'g_{\rho\rho}(\rho,\rho')\mu_{n}(\rho)\mu_{n}(\rho')
\end{equation}

\begin{figure}
\includegraphics[scale=0.64]{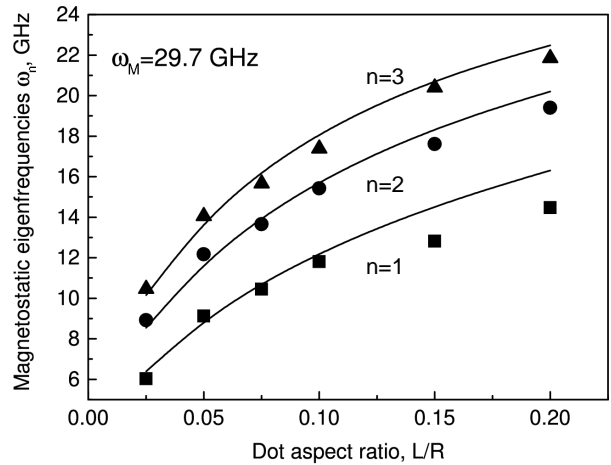}
\caption{\label{f_vfbeta}Dependence of the eigenfrequencies of the vortex 
radial modes with the indices n=1,2,3 on the dot aspect ratio $L/R$. The closed symbols correspond to numerical solutions of Eqs.~(\ref{Eq8}) and (\ref{Eq9}).}
\end{figure}

\begin{figure}
\includegraphics[scale=0.8]{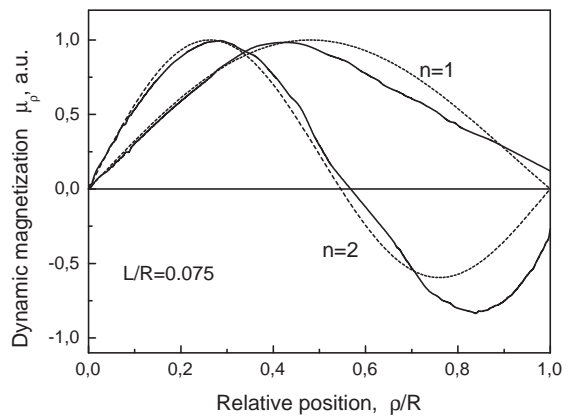}
\caption{\label{f_vef0-075}The vortex radial eigenfunctions $\mu _n(\rho )$ for the dot aspect ratio $\beta=0.075$ calculated by Eqs.~(\ref{Eq8}) and (\ref{Eq9}).
The solid lines are numerical solutions of Eqs.~(\ref{Eq8}) and (\ref{Eq9}),
the dashed lines are the Bessel functions $J_1(\kappa_n \rho)$.}
\end{figure}

\begin{figure}
\includegraphics[scale=0.8]{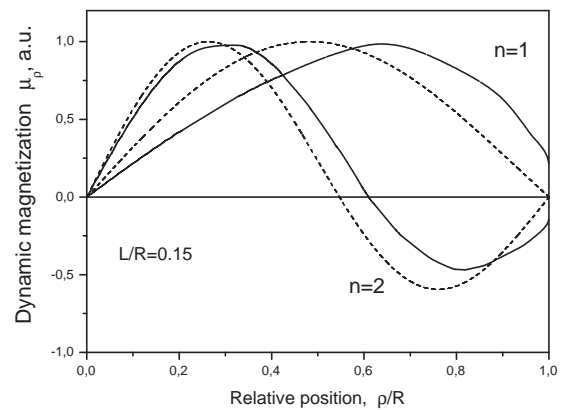}
\caption{\label{f_vef0-15}The vortex radial eigenfunctions $\mu _n(\rho )$ for the dot aspect ratio 
$\beta=0.15$. The solid lines are numerical solutions of Eqs.~(\ref{Eq8}) and (\ref{Eq9}),
the dashed lines are the Bessel functions $J_1(\kappa_n \rho)$.}
\end{figure}

In this magnetostatic approximation the vortex eigenfrequencies depend only on 
the dot aspect ratio $L/R$. More careful analysis shows that the 
eigenfrequencies are approximately proportional to $\sqrt{L/R}$ if
the aspect ratio $\beta = L/R \ll 1$.
The solution of the spectral problem Eq.~(\ref{Eq8})
will give us a discrete set of 
approximate magnetostatic eigenfunctions (radial modes profiles) and 
eigenfrequencies.  The kernel of Eq.~(\ref{Eq8}) is real and can be symmetrized.
This means that the Hilbert-Schmidt theory of the integral equations is 
applicable for our case. The eigenvalues are real
and the corresponding eigenfunctions are orthogonal.
The approximate numerical solution of the spectral problem (Eq.~(\ref{Eq8})) 
for FeNi dot
parameters (with $M_\mathrm{s}=800$~G, $\gamma/2\pi=2.95$~GHz/kOe and 
$\omega_M/2\pi=29.7$~GHz) and
different $L/R$ gives us the set of eigenfrequencies and eigenfunctions (see 
Fig.~\ref{f_vfbeta}, \ref{f_vef0-075}, and \ref{f_vef0-15}).
In the limiting case of a thin dot $\beta \ll 1$ this problem can be solved 
analytically.
We get eigenfrequencies $(\omega_n/ \omega_M)^2=f(\beta \alpha_n)$ and 
eigenfunctions 
$\mu_n(\rho)=C_n J_1(\kappa_n \rho)$, where $\alpha_n$ is the $n$th root of the
equation $J_1(x)=0$, $\kappa_n = \alpha_n /R$,  $J_1(x)$ is the first order Bessel function, and $C_n$ is the normalization constant. The set of $\kappa_n$ 
can be treated as a set of allowed quantized radial wave numbers, which is consistent with 
the geometry of a circular dot and its symmetry of the vortex ground state.
The equation for $\kappa_n$ corresponds to the strong pinning on the dot side 
surface $\rho=R$. In contrast, the numerical solution of Eq.~(\ref{Eq8}) corresponds to 
some intermediate values of the pinning,
although this pinning is strong enough. This pinning is of pure dipolar origin and it was
calculated for thin magnetic stripes in Ref.~\onlinecite{guslienko02_3}. The radial eigenfunctions $J_1(\kappa_n \rho)$ 
form a complete orthogonal
set and can be considered as good trial functions, which are asymptotically 
exact within the limit
$\beta \rightarrow 0$. The eigenfrequencies $\omega_n$ correspond to dynamic 
magnetostatic oscillations in the absence of the internal magnetic field.

\section{\label{s_mumag} Micromagnetic calculations and discussion}

\subsection{\label{s_femodel}Vortex static configuration}

The numerical 3D dynamic micromagnetic simulations are carried out using a 
hybrid finite element/boundary element method.\cite{fidler00} The nanodots are 
discretized into tetrahedral finite elements. The magnetic polarization is 
defined on the nodes of the finite element mesh and linear test functions are 
used on isoparametric elements. The effective field
which is the sum of the exchange, anisotropy, magnetostatic, and external field 
is calculated directly (for the local contributions from exchange, anisotropy, 
and external field) as $\vec{H}_\mathrm{eff} \approx -\partial W / \partial 
\vec{m}$ ($W$ is the energy density in Eq.~(\ref{Eq2})) and using a boundary element 
method\cite{fredkin90} for the long range magnetostatic field with open 
boundary conditions. The dynamic Landau-Lifshitz equation of motion for the 
magnetization is integrated with a preconditioned backward differentiation 
method.\cite{cohen96, suess02}
For the numerical micromagnetic finite element simulations we have assumed the 
material parameters of permalloy (Ni$_{80}$Fe$_{20}$), which is a typical soft 
magnetic material with negligible magnetocrystalline anisotropy.
We set the magnetic saturation polarization to
$J_{\mathrm{s}}=\mu_0 M_{\mathrm{s}} \approx 1~\unit{T}$
and the exchange constant to $A=13\ee{-12}~\unit{J/m}$, which gives an
exchange length $L_\mathrm{ex}=\sqrt{2A/M_\mathrm{s}^2}=5.7~\unit{nm}$.

\begin{figure}
\includegraphics[scale=0.4]{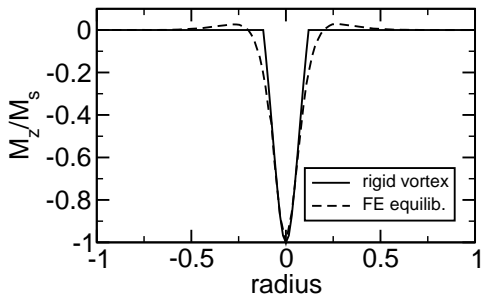}
\caption{\label{f_mzprof3}The calculated profile of the out-of-plane $M_z$ magnetization component in a circular nanodot ($R=200$~nm, $L=10$~nm). The solid line is solution of Eq.~(\ref{Eq0}), the dashed line corresponds to finite element micromagnetic simulations.}
\end{figure}

The finite element simulations have been initialized with the magnetization 
distribution of the vortex model based on Eq.~(\ref{eq_theta0}) and an 
approximate core radius. Then the Landau-Lifshitz equation of motion for the 
magnetization has been integrated with a damping constant $\alpha =1$ in zero 
field, and the magnetization relaxed to its equilibrium distribution, which 
minimizes the total Gibbs' free energy. The plots of the normal magnetization component $M_z$ in 
Fig.~\ref{f_mzprof3} show a comparison of the analytic vortex model with the 
equilibrium magnetization distribution of the FE simulation.

We define the vortex core radius by the equation $M_z (\rho)= 0$.
The numerical results show, that the vortex core is approximately 50\% larger (18.5~nm) 
than assumed by the vortex model given by Eq.~(\ref{eq_theta0}) (12~nm). Furthermore the finite element simulation 
shows that there is a region with $M_z < 0$ outside the core. Thus, we find 
positive surface charges in the core of the vortex, which are surrounded by 
negative surface charges. Only outside of approximately half the radius (50~nm) 
almost all surface charges disappear. It has been verified, that there is very 
little variation of the magnetization distribution across the thickness of the 
nanodot. The effect of small oscillations of the static vortex magnetization 
profile can be neglected for large $R>200$~nm.

The difference in magnetostatic and exchange energy between the approximate analytical vortex model given by Eq.~(\ref{eq_theta0}) and the finite element simulation is calculated. 
The static vortex magnetization distribution within the analytic vortex model has been obtained using the formulas 
given in Refs.~\onlinecite{usov93} and \onlinecite{usov94}. The equilibrium 
magnetization distribution after relaxation (dashed line in 
Fig.~\ref{f_mzprof3}) as calculated with the finite element model leads to 
a 4.4\% lower total energy. We 
confirmed numerically that the vortex core has relatively small radius $<20$~nm 
and, therefore, it can be ignored in the description of magnetization dynamics
of large dots with $0.2<R<2 \mu$m.\cite{scholz03_03,ivanov04}

\subsection{\label{s_translation}Vortex translation mode}

\begin{figure}
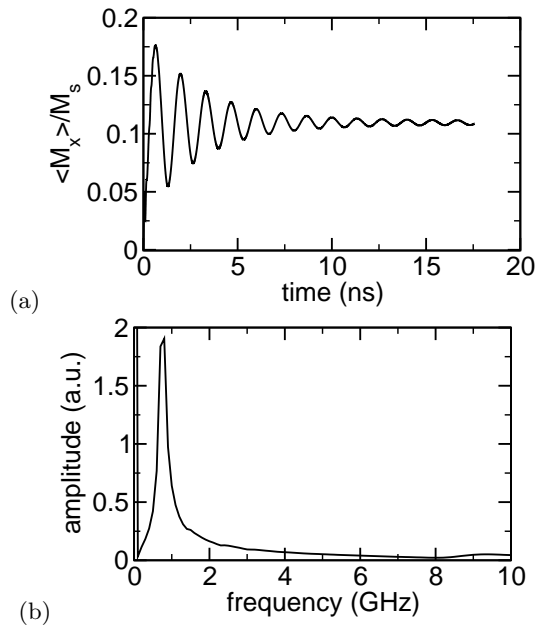

(a)
\includegraphics[scale=0.4]{fig5a.eps}
\\
(b)
\includegraphics[scale=0.4]{fig5b.eps}
\caption{\label{f_mxavg_t_0104}Oscillations of the in-plane averaged vortex magnetization component $\langle M_x
\rangle$ as a function of time (a) and its Fourier spectrum (b)
for the dot with $L/R=20~\unit{nm}/100~\unit{nm}=0.2$ under
applied in-plane field $H_x=0.01~\unit{T}$.}
\end{figure}

To study the vortex lowest frequency mode (translation mode) numerically an 
external field of 10~mT has been applied in the plane of the dot
to the remanent state and a damping 
constant of $\alpha=0.05$ has been assumed. The time evolution of $\langle M_x 
\rangle$ (the average of $M_x$ over the whole nanodot) for a dot with an aspect 
ratio of $L/R=10~\unit{nm}/100~\unit{nm}=0.1$ is given in 
Fig.~\ref{f_mxavg_t_0104}(a). Then the damped oscillation, which is caused by 
the spiral motion of the vortex core towards its equilibrium position, is 
observed. The corresponding Fourier spectrum is given in 
Fig.~\ref{f_mxavg_t_0104}(b) and shows a sharp peak at a frequency of 0.7~GHz.

\begin{figure}
 \includegraphics[scale=0.4]{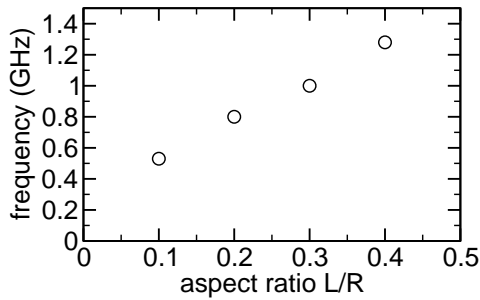}
\caption{\label{f_precess}Translation mode eigenfrequency versus the dot aspect ratio $L/R$ for cylindrical nanodots with $R=100~\unit{nm}$.}
\end{figure}

Fig.~\ref{f_precess} shows the results of calculations of the translation mode 
eigenfrequencies of various nanodots with a radius $R=100~\unit{nm}$ and a 
thickness between 10 and 40~nm. The translation eigenfrequency is approximately linearly proportional to the dot aspect ratio $L/R$. The results are in good agreement with the results of a finite difference model and the analytical ``two-vortices'' model presented in Ref.~\onlinecite{guslienko02}.

\begin{figure}
 \includegraphics[scale=0.4]{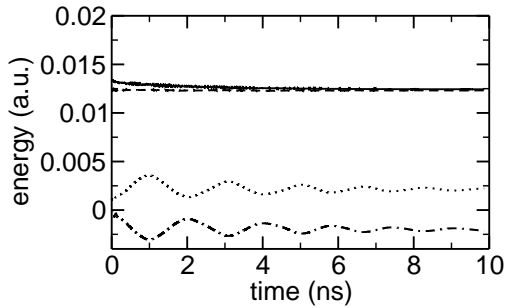}
\caption{\label{f_energy_0104}The contributions to magnetic energy vs. time for a cylindrical dot with $L/R=0.1$ and $H_x=0.01~\unit{T}$. Solid line: the total magnetic energy, dashed line: exchange energy, dotted line: magnetostatic energy, dot-dashed line: Zeeman energy.}
\end{figure}

The decreasing total energy (dissipation due to damping with $\alpha=0.05$ in 
the Landau-Lifshitz equation of motion) and the swapping between magnetostatic 
and Zeeman energy are shown in Fig.~\ref{f_energy_0104}. The exchange energy 
remains constant, because the vortex core, which accounts for most of the 
exchange energy, moves without changing its shape. This supports the analytical 
description of the translational mode suggested in 
Ref.~\onlinecite{guslienko02}, where circular motion of the vortex core as 
whole was calculated. Note that recently existence of this low-frequency vortex translation mode
was confirmed experimentally by direct Kerr microscopy measurements in Ref.~\onlinecite{park03}.
The eigenfrequency measured by Park et al.\ agrees with our calculations with an error of less than 20\%.

\subsection{\label{s_radial}Vortex radial modes}

The vortex radial modes were excited by applying an external field of 5~mT in 
$z$-direction and relaxing the magnetization with large damping ($\alpha=1$). 
When equilibrium was reached, the external field was instantaneously switched 
off and the free oscillation in zero field with very low damping was studied. 
With $\alpha=0.05$ the oscillations are damped out within a few oscillations. 
As a result the Fourier spectra are very poor because the resolution of the 
Fourier spectra increases with measurement time. Therefore $\alpha=0.001$ has 
been used to observe many oscillations and obtain a high resolution Fourier 
spectrum.

\begin{figure}
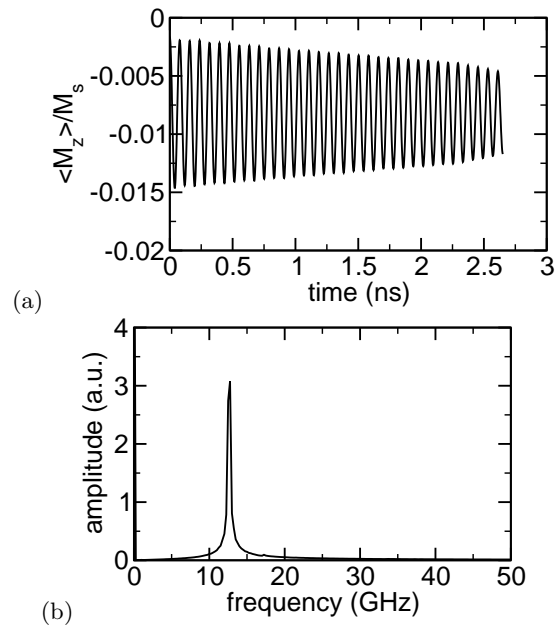

(a)
\includegraphics[scale=0.4]{fig8a.eps}
\\
(b)
\includegraphics[scale=0.4]{fig8b.eps}
\caption{\label{f_mzavg_t_020414} Oscillations of the averaged out-of-plane magnetization component $\langle M_z\rangle$ as a function of time (a) and its Fourier spectrum (b) for the cylindrical dot with $L/R=0.2$ under applied out-of-plane field
$H_z=0.005~\unit{T}$.}
\end{figure}

\begin{figure}
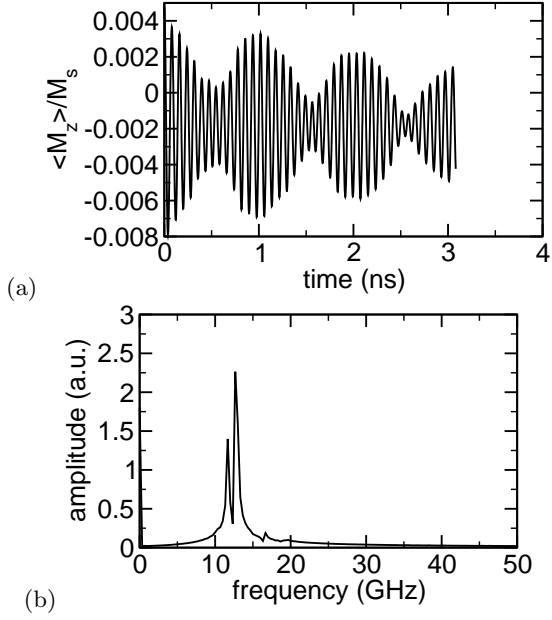

(a)
\includegraphics[scale=0.4]{fig9a.eps}
\\
(b)
\includegraphics[scale=0.4]{fig9b.eps}
\caption{\label{f_mzavg_t_020408} Oscillations of the averaged out-of-plane magnetization component $\langle M_z \rangle$ as a function of time (a) and its Fourier spectrum (b) for the cylindrical dot with $L/R=40~\unit{nm}/200~\unit{nm}=0.2$.}
\end{figure}

\begin{figure}
(a)
\includegraphics[scale=0.4]{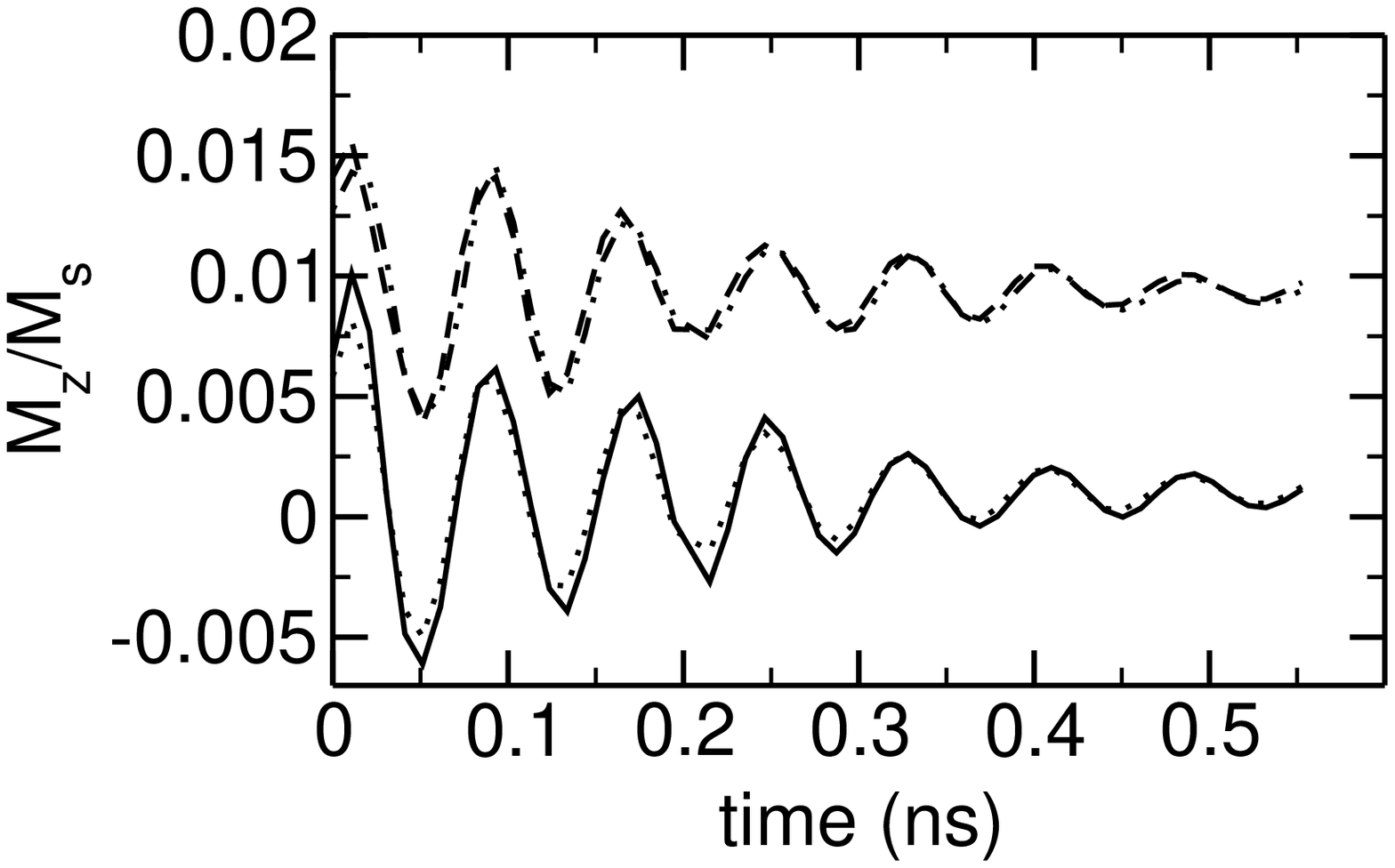}
\\
(b)
\includegraphics[scale=0.4]{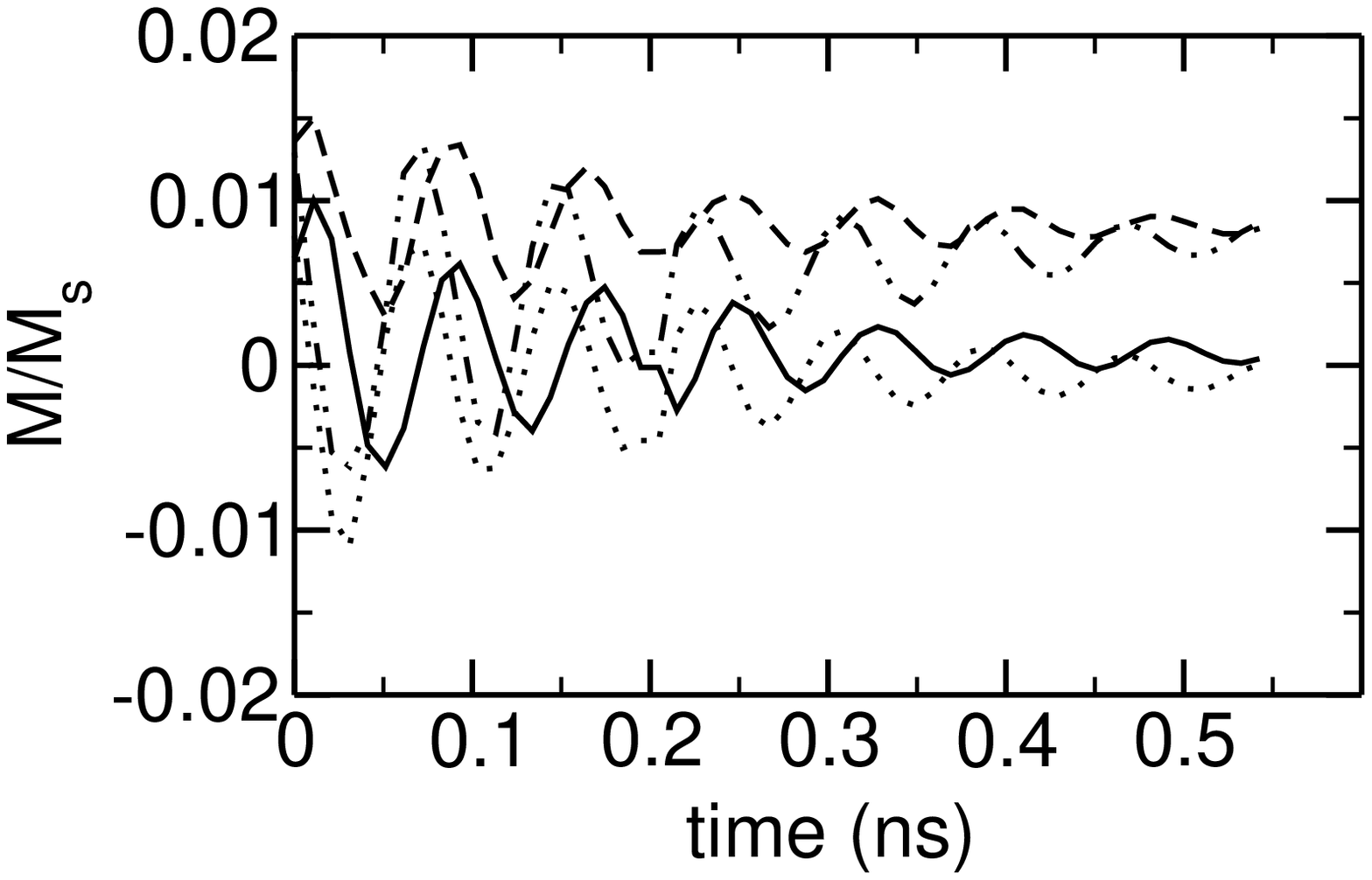}
\caption{\label{f_f_mrz_t}Oscillations of the vortex magnetization components. (a) $M_z$ (measured at different positions $\xi =x/R$,
solid line: $\xi =-1$, dotted line: $\xi =-0.8$, dashed line: $\xi =-0.5$, dot-dashed line: $\xi =+0.5$) oscillates in phase across the whole nanodot. The oscillation amplitude slightly decreases towards the center. (b) $M_{\rho}$ and $M_z$ as a function of time at different radii $r=\rho /R$. Solid line: $M_z$ at $r=1.0$, dotted line: $M_{\rho}$ at $r=1.0$, dashed line: $M_z$ at $r=0.5$, dot-dashed line: $M_{\rho}$ at $r=0.5$).
The phase shift between the $M_{\rho}$ and $M_z$ components is equal to 90~deg
(magnetization rotation in the $z$-$\rho$ plane).}
\end{figure}

\begin{figure}
\includegraphics[scale=0.5]{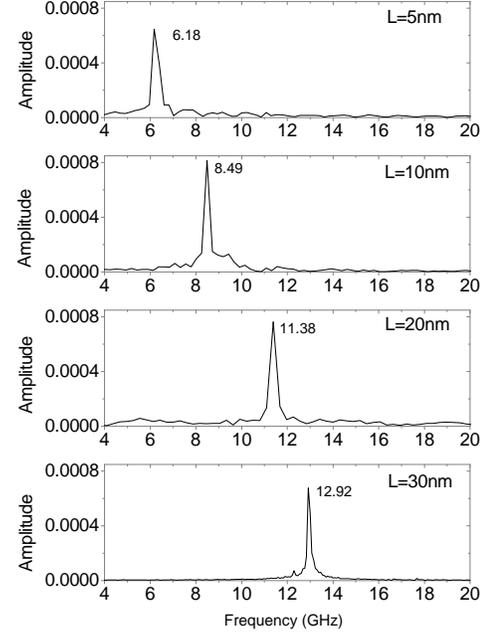}
\caption{\label{f_fourierevol} Evolution of the Fourier spectrum of the component $\langle m_z \rangle$ for the first ($n=1$) radial vortex excitation mode with varying the dot thickness $L$ from 5~nm to 30~nm. $R=200$~nm.}
\end{figure}

The time dependence of the average magnetization $\langle M_z \rangle$ and the 
Fourier spectra for nanodots with an aspect ratio of 
$L/R=20~\unit{nm}/100~\unit{nm}=0.2$ and $L/R=40~\unit{nm}/200~\unit{nm}=0.2$ 
are given in Figs.~\ref{f_mzavg_t_020414}(a) and (b) and 
\ref{f_mzavg_t_020408}(a) and (b), respectively. 
Note that the azimuthal modes (with the index $m \ne 0$) do not contribute to 
oscillations of $\langle M_z \rangle$. We checked numerically that the calculated modes 
have radial symmetry and correspond to the rotation of the dynamical magnetization 
$\vec{\mu}=(\mu_{\rho},0,\mu_{z})$ around the vortex static magnetization $\vec{m}_0=(0,m_0^{\phi},0)$
(cf. Fig.~\ref{f_f_mrz_t}).

For the constant aspect ratio $L/R=0.2$ we find an eigenfrequency of approximately 
12.6~GHz. However, for a nanodot with $L/R=40~\unit{nm}/200~\unit{nm}=0.2$ a 
very pronounced beating is observed (Fig.~\ref{f_mzavg_t_020408}). This is due 
to the fact, that there is another eigenfrequency of 11.6~GHz very close to the 
12.6~GHz oscillation.
However, the main peak position depends only on the combination ratio $L/R$. 
This confirms the magnetostatic origin of the mode. But the physical picture is 
more complicated for large $L$, when the magnetization dependence on $z$ may be 
essential.

\begin{figure}
 \includegraphics[scale=0.4]{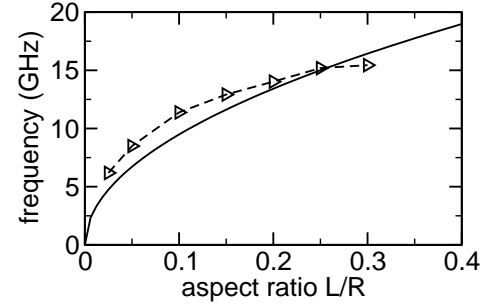}
\caption{\label{f_radmode}The radial vortex mode eigenfrequencies versus aspect ratio $L/R$ for nanodots with different thickness $L$. The open triangles are points simulated numerically, the solid line corresponds to fitting using the equation $\omega_n \sim \sqrt{L/R}$.}
\end{figure}

We concentrated on numerical simulations of the dependence of the first magnetostatic radial eigenfrequency on the dot aspect ratio $L/R$. 
The calculated evolution of the first radial mode with increasing dot thickness $L$ (5-30~nm) is shown in Fig.~\ref{f_fourierevol}. The eigenfrequency of this mode increases with $L$ in agreement with analytical calculations of the vortex 
magnetostatic modes in Sec.~\ref{s_theory}. We plotted the micromagnetically simulated eigenfrequency as a function of the dot aspect ratio $L/R$ in Fig.\ref{f_radmode} together with the analytic dependence $\omega_n \sim \sqrt{L/R}$ and got a good agreement of the eigenfrequencies calculated by these different methods. Thus, our numerical calculations confirmed that the radial eigenfunctions and eigenfrequencies which correspond to low-lying part of the vortex dot excitation spectrum have magnetostatic origin.

The calculated radial eigenmodes are magnetostatic (wave length is of the order of the dot radius) and
concentrated in the area outside of the vortex core (see 
Figs.~\ref{f_vef0-075}, \ref{f_vef0-15} for the calculated eigenfunctions). The 
eigenfrequencies of these modes 
$\omega _n$ are determined by the dynamic demagnetizing fields. The static 
dipolar field for the vortex structure has two components $H_z$ and $H_{\rho}$, 
which depend on $\rho$ only due to circular symmetry of the vortex.
Both components are negligibly small in the area of our interest, i.e.\ 
outside the vortex core.
In this sense the problem of non-uniform magnetostatic spin-excitations is 
similar to the problem of magnetostatic
waves in thin magnetic stripes magnetized along their long 
side considered by Guslienko et al.\ in Ref.~\onlinecite{guslienko02_3}.
Effective boundary conditions for dynamic magnetization 
on the dot surface side can be also formulated similarly to Ref.~\onlinecite{guslienko02_3}. The condition $\mu_n(\rho 
=0)=0$ is satisfied near the dot center. This is consistent with our assumption 
about concentration of the dynamic magnetization of the radial modes outside 
the vortex core. These radially symmetric magnetic dynamic excitations can be 
detected by modern experimental techniques such as BLS, FMR,and time-resolved Kerr- and magnetic circular dichroism measurements.

Very recently, when this manuscript was ready for submission, we became aware
of experimental data by Buess et al.\cite{buess04}
They measured the eigenfrequencies of the vortex radial modes
by time-resolved Kerr microscopy and obtained frequencies of
2.8, 3.9 and 4.5~GHz. The dot aspect ratio $\beta=L/R=15~\unit{nm}/3000~\unit{nm}=
0.005$ is very
small and we can apply the approximate equation $(\omega_n/\omega_M)^2=
f(\beta \alpha_n)$ derived from Eq.~(\ref{Eq_wn_wM}).
The calculated eigenfrequencies 2.91, 3.93 and
4.73~GHz are very close to the experimental data.

\section{\label{s_conclusions}Conclusions}

The vortex eigenfrequencies and magnetization profiles were calculated on the 
basis of analytical and numerical approaches.
The low-lying part of the spectrum of spin excitation over the vortex ground 
state consists of discrete eigenfrequencies
of magnetostatic origin. The corresponding dynamic eigenmodes are localized 
near the dot center (the lowest translation mode)
or are excited mainly outside the vortex core (radially symmetric magnetostatic 
modes).
The model of the shifted vortex with no side surface charges\cite{guslienko02}
well explains the results of our micromagnetic numerical calculations of the 
translation
mode eigenfrequency. The set of magnetostatic eigenfunctions $\mu _n(\rho)$ 
and eigenfrequencies $\omega_n$ was calculated as a function of the dot aspect 
ratio $L/R$. The first magnetostatic eigenfunction can be approximately 
described as a uniform mode (no nodes). Whereas the high frequency 
eigenfunctions have a number of the nodes proportional to their number. The 
frequencies of the radial eigenmodes are proportional to $\sqrt{L/R}$ for thin 
magnetic dots with dot thickness $L \approx L_\mathrm{ex}$ and dot radii
of about 1~$\mu$m.

\begin{acknowledgments}
Work at ANL was supported by US DOE BES grant \#W-31-109-ENG-38.
The authors thank C.~A.~Ross for sending her manuscript prior to its 
publication and M.~ Grimsditch for discussion of our manuscript.
\end{acknowledgments}

\bibliography{guslienko}

The submitted manuscript has been created by the University of Chicago as Operator of Argonne National Laboratory (``Argonne'') under Contract No. W-31-109-ENG-38 with the U.S. Department of Energy. The U.S. Government retains for itself, and others acting on its behalf, a paid-up, nonexclusive, irrevocable worldwide license in said article to reproduce, prepare derivative works, distribute copies to the public, and perform publicly and display publicly, by or on behalf of the Government.

\end{document}